\documentclass[10pt,preprint]{aastex}

\newcommand{\numberCatalogEBs}{2605}
\newcommand{\numberShortEBs}{1279} % (in_cat and short)
\newcommand{\numberTM}{236}
\newcommand{\rateTriple}{$\sim$ 20\%}
\newcommand{\numberTertiaryEclipse}{3}
\newcommand{\rateTertiaryEclipse}{$<$ 1\%}
\newcommand{\numberTMsectionAB}{111} %AB but not tertiary (tablenote 5)
\newcommand{\rateTMsectionAB}{$\sim$ 10\%} % above / numberShortEBs
\newcommand{\numberTMsectionC}{94} %C but not tertiary
\newcommand{\rateTMsectionC}{$\sim$ 7\%} % above / numberTM

\newcommand{\numberDV}{Nine} % DV and TM % 
\newcommand{\DVKICs}{3936357, 4069063, 5310387, 6629588, 7375612, 8122124, 8758716, 10014830, and 10855535}
\newcommand{\numberTT}{Thirty-one}

\usepackage{float}
\usepackage{amsmath}
\usepackage[normalem]{ulem}

\shorttitle{ETVs for Kepler close EBs}
\shortauthors{Conroy et al. 2013}

\begin{document}

\title{\emph{Kepler} Eclipsing Binary Stars. IV. Precise Eclipse Times for Close Binaries and Identification of Candidate Three-Body Systems}
\author{Kyle E. Conroy}
\affil{Department of Physics and Astronomy, Vanderbilt University, VU Station B 1807, Nashville, TN 37235}
\affil{Department of Astrophysics and Planetary Sciences, Villanova University, 800 E Lancaster Ave, Villanova, PA 19085}

\author{Andrej Pr{\v s}a}
\affil{Department of Astrophysics and Planetary Sciences, Villanova University, 800 E Lancaster Ave, Villanova, PA 19085}

\author{Keivan G. Stassun}
\affil{Department of Physics and Astronomy, Vanderbilt University, VU Station B 1807, Nashville, TN 37235}
\affil{Department of Physics, Fisk University, Nashville, TN 37208}

\author{Jerome A. Orosz}
\affil{Department of Astronomy, San Diego State University, 5500 Campanile Drive, San Diego, CA 92182}

\author{Daniel C. Fabrycky}
\affil{Department of Astronomy and Astrophysics, University of Chicago, 5640 S. Ellis Ave., Chicago IL 60637}

\author{William F. Welsh}
\affil{Department of Astronomy, San Diego State University, 5500 Campanile Drive, San Diego, CA 92182}

\begin{abstract}

We present a catalog of precise eclipse times and analysis of third body signals among \numberShortEBs{} close binaries in the latest \emph{Kepler} Eclipsing Binary Catalog.
For these short period binaries, \emph{Kepler}'s 30 minute exposure time causes significant smearing of light curves.  
In addition, common astrophysical phenomena such as chromospheric activity, as well as imperfections in the light curve detrending process, can create systematic artifacts that may produce fictitious signals in the eclipse timings.
We present a method to measure precise eclipse times in the presence of distorted light curves, such as in contact and near-contact binaries which exhibit continuously changing light levels in and out of eclipse.
\numberTM{} systems for which we find a timing variation signal compatible with the presence of a third body are identified.
These are modeled for the light time travel effect and the basic properties of the third body are derived.
This study complements \citet{KepEBetv1}, which focuses on eclipse timing variations of longer period binaries with flat out-of-eclipse regions.
Together, these two papers provide comprehensive eclipse timings for all binaries in the \emph{Kepler} Eclipsing Binary Catalog, as an ongoing resource freely accessible online to the community.

\end{abstract}

\section{Introduction}

Eclipsing binaries have historically contributed a wealth to stellar astrophysics.
They have been used to determine distances, compute fundamental stellar parameters, and test stellar evolution models.
The \emph{Kepler} mission \citep{Borucki,Batalha} and its unprecedented precise photometry of $\sim 160,000$ stars, has allowed us to create a catalog of \numberCatalogEBs{} eclipsing binaries (hereafter EBs) in the \emph{Kepler} field \citep{KepEB4,KepEB2,KepEB1}.
This catalog is rich in interesting objects for individual study and also presents a large sample of EBs for statistical analysis.
In studying this sample, we can attempt to determine the occurrence rate of EBs, circumbinary planets, and multiple star systems.

Some theories for short-period binary star formation call for the presence of a third-body.  
In these scenarios, the close binary was not created \emph{in situ}, but rather at a larger separation as a part of a multiple star system \citep{Bonnell}.
Tidal friction and Kozai cycles between the inner-binary and a companion can cause the inner-orbit to shrink over time \citep{FabTrem}, and result in a hierarchical multiple system \citep{Reipurth}.
The spectroscopy and imaging studies by \citet{Tok97} and \citet{Tok06} have found 40\% of binaries with periods less than 10 days, and 96\% with periods less than 3 days, have a wide tertiary companion.
The general interpretation of these findings is that the tightest binaries likely became hardened over time through interactions with the tertiary companion, and the system evolves toward an increasingly hierarchical
configuration. Indeed, the SLoWPoKES study of ultra-wide binaries in the Sloan Digital Sky Survey \citep{slowpokes} found that the widest visual pairs with physical
separations of 0.01--1 pc, in fact contain a tight binary $\sim$80\% of the time \citep{law}, again confirming the general picture that tight binaries are nearly always accompanied by wide tertiaries and that the tightest binaries are accompanied by the widest tertiaries.

Discovery and study of these multiple systems gives new insight into the physics of EBs.
Statistically, we can compare observed rates of multiple systems to theoretical models for short-period binary formation.
We can also model each system individually to study the disruptive dynamical effects seen in some cases.

\citet{KepEB4} determines ephemerides for the entire \emph{Kepler} Eclipsing Binary Catalog.
If there are no external effects, a linear ephemeris will correctly predict all eclipse times of an EB.
By measuring the exact time of each eclipse for a particular binary and comparing it to the calculated time from the linear ephemeris, we can create an ETV curve (`eclipse timing variations'; sometimes also referred to as an O-C diagram).
Any trend in these timing residuals may be the result of one or more physical effects occurring in the system. 

Using transit timings and eclipse timings to find exoplanets is a well-known method \citep{Schwarz}.  
\citet{Fab}, \citet{Ford} and \citet{Steffen} used transit timings to detect and study multiple planetary systems, 
while \emph{Kepler} 16 \citep{Kep16}, 34, and 35 \citep{Kep34} were validated, in part, through their eclipse timing variations.
The processes that can induce ETVs, which are the focus of this paper, include the following:

\begin{itemize}
\item Light Time Travel Effect (LTTE): a third-body perturbing the center of mass of the binary system creates a light-time delay along the line of sight which can cause eclipses to appear earlier or later than expected.
\item Non-hierarchical third-body: the presence of a third-body actually changes the period of the binary over time.
\item Mass transfer: mass transfer between the components in the binary changes the period.
\item Gravitational Quadrupole Coupling (Applegate effect): spin-orbit transfer of angular momentum in a close binary due to one of the stars being active produces period changes up to $10^{-5}$ times the binary period \citep{Applegate}.
\item Apsidal Motion: the rotation of the line of apsides causes a change in the time between primary and secondary eclipses even though the period remains unchanged (requires an eccentric orbit).
\item Spurious Signals: due to spots and other effects that distort the EB light curve.
\end{itemize}

\citet{Rapp} previously published a list of 39 candidate third-body \emph{Kepler} systems using eclipse times and \citet{Gies} published a preliminary study on timing variations in 41 \emph{Kepler} Eclipsing Binaries.
\citet{KepEBetv1} will provide eclipse times for detached binaries, and this paper provides eclipse times for close binaries.
Together, these two papers will comprehensively cover all \numberCatalogEBs{} binaries in the catalog.

\emph{Kepler}'s essentially uninterrupted observing over a long time baseline presents the opportunity to precisely time the eclipses and detect any underlying signals due to third bodies, apsidal motion, dynamical interaction, etc.
Due to the large number of EBs in the entire catalog, it is necessary to create an automated method for timing eclipses across the catalog.
Short period and overcontact systems present a particular challenge due to spot activity and data convolution, due to a relatively long integration time.

In this paper we discuss our method for automating eclipse timings for close \emph{Kepler} EBs in Section 2.
Eclipse timings are reported for \numberShortEBs{} binaries in the catalog in Section 3.
In Section 4, light time travel effect models for the \numberTM{} that are flagged as potential third-body candidates are also provided.
We discuss our findings in Section 5 in the context of binary formation and evolution theory, and summarize our conclusions as well as
information for accessing the products of our comprehensive eclipsing timing measurements in Section 6.

\section{Data and Methods}
\subsection{Sample of Eclipsing Binaries}
\citet{KepEB4} will update the \emph{Kepler} Eclipsing Binary Catalog, raising the count of EBs from 2165 to \numberCatalogEBs{}.
The database is kept up-to-date with future data and revisions at \texttt{http://keplerEBs.villanova.edu}.
As changes and updates are made to the catalog, ETVs are being recomputed and updated automatically and made available in real-time through the online catalog.

\begin{figure}[h]
\plotone{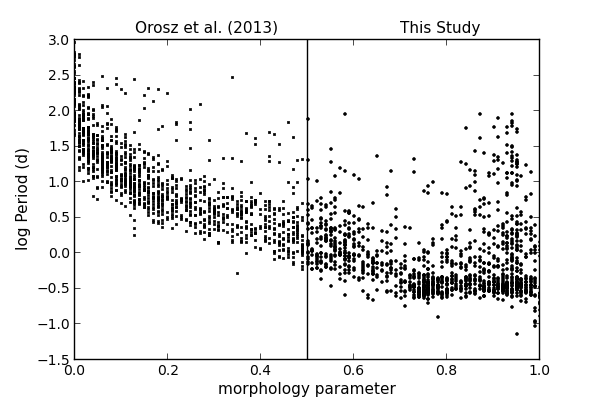}
\caption{Period vs morphology parameter for the binaries in the \emph{Kepler} Eclipsing Binary catalog.  Objects included in this paper have a morphology parameter greater than 0.5.}
\label{etvcat}
\end{figure}

\citet{KepEBetv1} will provide eclipse times for binaries with flat out-of-eclipse regions, covering most of the detached binaries with periods greater than 1 day.
There we locally detrend each eclipse and use a piecewise Hermite spline template to determine the time of mid-eclipse.
This technique performs well on the set of detached systems but is not optimal for overcontact systems, systems with strong reflection effects or tidal distortion, or short-period binaries with only a few points in each eclipse due to \emph{Kepler}'s 30 minute cadence.
For this reason, we divide the catalog based on the morphology parameter as described in \citet{KepEB3}.  
This parameter is a value between 0 and 1 which describes the ``detachedness'' of an eclipsing binary, with 0 being completely detached and 1 being overcontact or ellipsoidal.
\citet{KepEBetv1} report timings for binaries with a morphology parameter less than 0.5.
Our method addresses and determines eclipse times for the remainder of the \emph{Kepler} Eclipsing Binary Catalog.  
The distribution of the catalog between these two methods is shown in Fig.~\ref{etvcat}, with \numberShortEBs{} binaries in the sample for this paper.

\subsection{Light Curve Preparation}

We detrend and phase ``SAP'' (simple aperture photometry) \emph{Kepler} data through Q16 as described in \citet{KepEB1}.  
The upper envelope of the raw data is fit with a chain of Legendre polynomials using a sigma-clipping technique and manually setting the breaks between sections and orders of the polynomials.
The data are then divided by this fit, resulting in a flat baseline.  
These detrended data are then phased on the linear ephemeris as reported in \citet{KepEB4}, and used as input into the ETV code, described below.

\subsection{Measuring Eclipse Times}
\begin{figure}[h]
\plotone{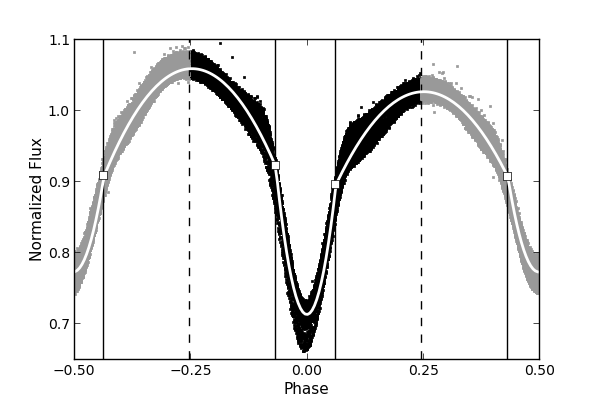}
\caption{Typical polyfit and eclipse bounds for a semi-detached binary.  
The polyfit knots are indicated with the squares and solid vertical lines, with the polyfit drawn in white over the data.  
Data considered as part of the primary eclipse are shown in black while those belonging to the secondary eclipse are shown in gray. 
The eclipse bounds are set at the arithmetic bisector of the adjacent knots and are shown with dashed vertical lines.}
\label{ecl_bounds}
\end{figure}

We fit a polynomial chain to the phased light curve data as described in \citet{Prsa08}.  
This analytic function is a chain of four polynomials that is continuous, but not necessarily differentiable, at knots which were optimized to find the best overall solution.
This function does not represent a physical model, but rather analytically describes the mean phased shape of the binary light curve, an example of which can be seen in Fig.~\ref{ecl_bounds}. 

\begin{figure}[h]
\plotone{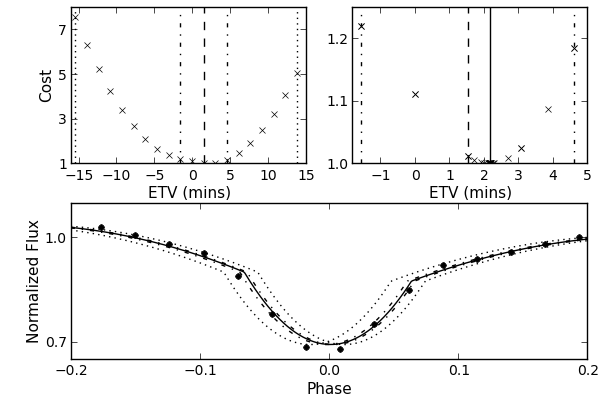}
\caption{Reduced cost function ($\chi^2$) values, shown as x's, are computed heuristically (top-left) for 20 evenly-spaced phase-shifts within 0.05 phase, shown by the dotted lines in all panels.  The best fit of these is shown with the dashed line in the top-left panel.  A bisection approach (top-right) is then applied in the area surrounding this estimate, as shown by the dot-dashed lines.  This results in a final minimum at the phase shift denoted by the solid line.  The bottom plot shows the data for a single eclipse along with the polyfits for the respective shifts noted above.}
\label{bisection}
\end{figure}

We then take this analytical representation and, using a combination of heuristic and bisection approaches, determine the horizontal shift required to minimize the $ \chi^2 $ (cost function) for each individual eclipse as shown in Fig.~\ref{bisection}.
In order to minimize the effect due to spots or imperfect detrending, a vertical shift is first determined using linear least squares for each eclipse and is applied before computing cost functions for horizontal shifts.
The cost function is initially sampled at 20 evenly-spaced phase shifts between -0.05 and 0.05 phase.
The minimum of this sampling is then used as the center of the bisection algorithm to quickly find the local minimum of the cost function.
The resulting $\chi^2$ values are unusually large because the errors on the \emph{Kepler} data are only formal and do not include any absolute calibration effect \citep{Jenkins}.  
Therefore, for each eclipse, we normalize the entire cost function such that the minimum cost is set to $N-p-1$, where $N$ is the number of data points used for that eclipse and $p$ is the degrees of freedom, which we take to be 1.
This reduced cost function is then used to compute 1-sigma errors on each timing to correspond to the $\Delta \chi^2 = 1$ contour.

For the shortest binaries in the catalog, however, the long-cadence data result in significant phase-smearing and limits our method to a very minimal number of points per cycle to determine a fit.
If there were to be a third-body, the signal would likely be buried in the noise induced by these factors.
For this reason, we include as many data points as possible in each eclipse timing.
Each data point is considered to belong to an eclipse if its phase as determined by the initial linear ephemeris is within bounds.
We initially set these bounds to be the mid-point between polyfit knots in the out-of-eclipse region as shown in Fig.~\ref{ecl_bounds}.
To improve results for particular objects being studied individually, changing these bounds to use the knots (instead of the mid-points) can sometimes lower the systematics in the signal.
For any given eclipse, if the region between these bounds is not fully sampled or does not have at least 3 data points, then timings are not computed for that eclipse.
Eclipse timings are then compared to the values expected from the linear ephemeris as reported by \citet{KepEB4} to compute the residuals and test for the presence of an ETV signal.  

\subsection{Dealing with Sources of Spurious ETV Signals}

Due to a typically small number of points per eclipse, our timings are sensitive to various imperfections in the data processing, affecting the measured eclipse time and potentially introducing noise and/or fictitious signals in the ETV signal.
Instrumental or astrophysical pulsations on top of the binary signal can change the shape of a single eclipse which can mimic a timing variation.
The detrending process attempts to remove these additional signals, but is not perfect, struggles at removing signals that happen during eclipse, and can also introduce spurious signals.
Also, all polyfits in the current version of the catalog use chains of four second-order polynomials, which does not always result in the ideal fit and can leave slight phase-dependent residuals.
For the purpose of pipeline processing, we limit ourselves to second order polynomials, but note that, for special cases and in-depth studies, higher precision timings can be obtained by increasing the order of the fit.
Until all polyfits are updated to a higher order in the future, we will use the second-order fits and manually run ETV signals with a higher order for any individual ETV signals that warrant further study.
In the cases when a binary has a period that is near-commensurate to \emph{Kepler}'s 29.44 minute cadence, the period and cadence may beat, which results in a separate spurious signal.
Any combination of these effects can cause issues in determining true and precise eclipse times when dealing with only a few data points.

\begin{figure}[h]
\plotone{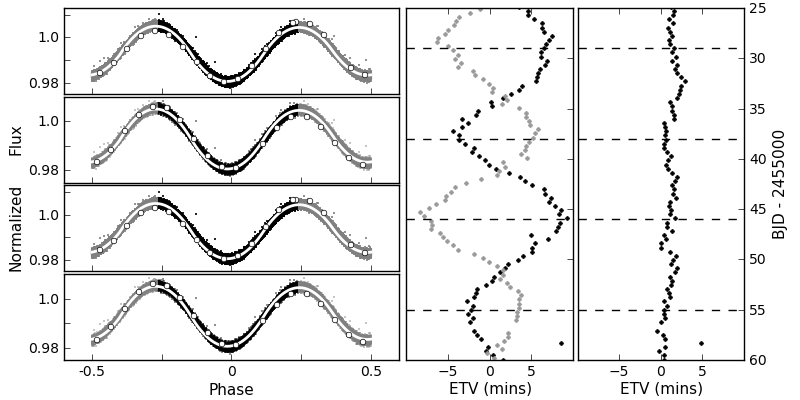}
\caption{Determining eclipse timings using both eclipses will cancel the anti-phase effect and reveal any underlying signal.  
The plots on the left show a phased light curve with primary eclipse data in black and secondary in gray with the analytic `polyfit' in white.
These four plots highlight the data during individual cycles (shown in white on the left) at four different times noted in the ETV plots with the dashed line, showing the presence of spots.
The plots on the right show the ETV as measured at for primary and secondary eclipses separately (middle) and full phase (right).
}
\label{anti-phase}
\end{figure}

\begin{figure}[h]
\plotone{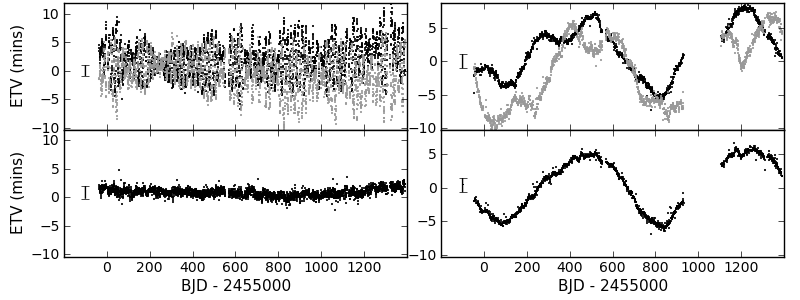}
\caption{ETVs for KIC 6880727 (left) and 4451148 (right) determined for primary and secondary eclipses separately (top) and together (bottom).
KIC 6880727 (left) shows an example with no underlying signal under the antiphase ``noise'', while KIC 4451148 (right) shows a possible underlying third-body signal.
Typical errors for ETV measurements are shown to the left of the data.}
\label{anti-phase-beforeafter}
\end{figure}

Fig.~\ref{anti-phase} demonstrates how the cost function for the phase shift is affected by the vertical discrepancy in the out-of-eclipse region, creating a fictitious signal in which the ETVs of the primary and secondary eclipse are in anti-phase.
The left of Fig.~\ref{anti-phase} plots four different eclipses, showing that over time the data in the out-of-eclipse region can be higher on either the right or the left.
When measuring timings for the primary and secondary separately, the cost function will artificially be minimized by ``pulling'' the analytic function towards the region with lower flux.
Since this will affect the primary and secondary in the opposite direction, we can mitigate for this effect by also running the fit over the entire phase.
This effectively averages out the anti-phase effect in the primary and secondary eclipses, projecting the real ETV signal of the entire system.
Fig.~\ref{anti-phase-beforeafter} shows two cases where the anti-phase signal was removed, clearly showing whether there is a presence of any underlying ETV signal.
These signals that show a ``random walk'' nature are discussed by \citet{Tran}.

Unfortunately, since there is no rigorous way to discriminate between true and fictitious anti-phase signals, this process would also hide a physical ETV signal such as apsidal motion.
Since we are dealing with short-period binaries, most of these orbits will be quite circular so we do not expect to be able to detect any systems with apsidal motion anyway.

\subsection{Short-Cadence Data}

If a binary has short-cadence data available, they are usually limited to a short time baseline.  
Since we are generally looking for long period trends in the ETV signal, we measure timings from the full long cadence dataset.
We also run timings on any available short cadence data in case some signal can be detected.

A few short period binaries in the catalog that appear to be overcontact or ellipsoidal variables could actually be detached systems whose light curves are convolved by \emph{Kepler}'s 30 minute long cadence exposure.
Since these systems are less prone to timing noise due to spots or mass transfer than true overcontact systems, the phase smearing and limited data points per eclipse are the main issues preventing us from recovering any ETV signal.
For this reason, short cadence data were requested and obtained via Director's Discretionary Time for 31 short-period detached EBs in the catalog without previous short-cadence observations in the hope of detecting third-body ETV signatures which were not visible in the long-cadence data.
Unfortunately, none seem to exhibit any significant ETVs.

Eclipse times are computed for all available short-cadence data, but due to the longer baseline, long-cadence timings are used for detection and fitting of potential third-body orbits.

\section{Results}
\subsection{Precise Eclipse Times}
Eclipse timing variations on the individual eclipses and the entire phase have been run for all objects with a morphology parameter greater (less detached) than 0.5 in the latest \emph{Kepler} Eclipsing Binary Catalog.
Our method requires at least 3 data points per timing, which allows us to get primary and secondary eclipse timings individually for long cadence data of binaries with periods as short as 3 hours and full phase timings for binaries as short as 1.5 hours.
We are able to determine timings for any binary with short cadence data, but since short cadence data availability is sparse and generally not for the whole length of the mission, short cadence ETVs are determined separately.

Plots and data for detrended light curves and eclipse times for the entire sample are available as a part of the \emph{Kepler} Eclipsing Binary online catalog at \texttt{http://keplerEBs.villanova.edu}.
An excerpt of the eclipse times is shown in Table \ref{tableetvs}.
With the third version of the catalog being released shortly, the database will be updated as new data become available and ephemerides are further refined.
This ETV code is incorporated into the pipeline: as objects are updated or added, their ETVs are recomputed and updated in real-time.

As ETVs are computed, the ephemerides in the catalog are refined by fitting a linear fit through the entire-phase timings and adjusting the values as necessary to get a ``flat'' trend.
For any ETV with a long-term sinusoidal trend, this could introduce systematics depending on the part of the sine curve observed and used to fit the linear trend.
In particular, for very long ETV signals (of the order of 1000 days and more), the measured orbital period of the binary will be anomalous because the variation cannot be accounted for from available data.

\subsection{Causes of an ETV Signal}

All ETV measurements were examined by eye for the presence of any interesting signal, discarding any that seem to be spurious based on their individual primary and secondary eclipse timings.
We do not expect to see evidence of apsidal motion in many of our targets due to their short periods and, consequently, circular orbits.
We also do not expect to be able to detect any signals due to gravitational quadrupole coupling.  
This mechanism is able to create period changes with amplitudes on the order of $10^{-5}$ times the period of the binary, meaning a maximum of 3.5 seconds for a binary with a period of 4 days, falling well within our noise limits.

ETV signals that are sinusoidal in nature or show any sign of curvature are flagged and fit for both a third-body signal and a parabolic mass transfer model.
For the cases where we only see a sign of curvature and not a full cycle, we could either be seeing a section in a long period third body signal or mass transfer.
To determine whether we consider the signal as a candidate third body or mass transfer, we compare the two models using the Bayesian Information Criterion \citep{BIC}:
\begin{equation}
	BIC = n \ln \left( \frac{1}{n} \sum \left( x_i - \hat{x_i} \right)^2\right) + k \ln n
\end{equation}
where $x_i$ are the data, $\hat{x_i}$ the model, $n$ the number of data points, and $k$ is the number of parameters used in the fit.
In the case of the eccentric LTTE model $k=6$, for the circular LTTE model $k=4$, and for the mass transfer model $k=3$.
The better fit, as determined by the lower BIC value, then determines whether we consider the signal as a candidate third body or mass transfer.

\subsection{ETVs with Parabolic Signals}

\begin{figure}
\plotone{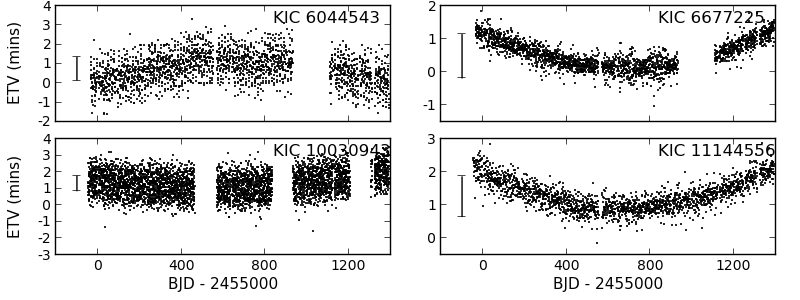}
\caption{A selection of ETV signals that are better fit by a quadratic ephemeric than a LTTE fit.}
\label{etv_MT}
\end{figure}

\numberTT{} ETV signals were better fit by a parabola than an LTTE orbit, and are possibly caused by mass transfer or the Applegate effect instead of the presence of a third body \citep{Hilditch}.
A selection of these signals are shown in Fig.~\ref{etv_MT}, all KICs are listed in Table \ref{tableMT}, all of which are available on the online catalog.

\subsection{ETVs with Third-Body Signals}

\numberTM{} binaries (\rateTriple{} of the sample) were flagged as candidate third bodies.
The results of the model fits are reported in Table \ref{tablerapp} with a selection plotted in Fig.~\ref{etv_LTTE}.
Based on the fitted period, we then divided these third body candidates into three sections.
The first group contains third body signals with periods less than 700 days, such that there are at least two full cycles of the signal present in data through Q16.
These systems have the highest confidence and are most likely due to the presence of a tertiary component.
The second group contains signals with periods between 700 and 1400 days, such that there is at least one full cycle present.
The last group contains signals with periods longer than 1400 days.  
Often these detections merely show some sign of curvature in the ETV signal and so a full sinusoidal signal cannot yet be confirmed.
For this reason the fits generally have large errors and many of these may not even be true triple systems, particularly the signals on the closest binaries which are more likely to be due to mass transfer.

\citet{Gies} presented an initial study of eclipse timings in 41 \emph{Kepler} binaries.
Of their entire sample of 41 binaries, 40 are still in the \emph{Kepler} EB Catalog (KIC 4678873 has since been removed from the catalog as a false positive), 32 fall under the scope of this paper (have a morphology less detached than 0.5), and 9 appear in our list of third-body signals.
They identified 14 out of their original 41 as candidate third-body systems, with others being identified as likely caused by starspots, pulsations, and apsidal motion.
Of their 14 candidate third-body systems, all 14 are still in the \emph{Kepler} EB Catalog, 12 fall under the scope of this paper, and 9 appear in our list of third-body signals.
Those that appear in both lists are noted in Table \ref{tableresults}.
Binaries that they list as candidate third-body systems, but we do not, either show significant noise or would be very long period LTTE orbits. 

\citet{Rapp} recently reported 39 triple-star \emph{Kepler} binaries due to ETV signatures.  
Of these 39, 21 fall under the scope of this paper, 19 of which also appear in our list of third-body signals, with the other two determined to be unlikely caused by a third-body due to their very short periods and notable spot activity.
The detections that overlap both of these studies are also noted in Table \ref{tablerapp}.
In most cases, the model fits from both studies are consistent.
In general, due to our treatment of the full \emph{Kepler} dataset now available, tertiary parameters should now be more precise and longer period third body signals are now more apparent.
Any disagreement is likely due to a slightly differ inner-binary ephemeris or the addition of the physical delay in their models (discussed further below).

\section{Analysis of Third Body Signals}

\subsection{Light Time Travel Effect Analysis}

\citet{Bork} presents analytic functions for the light time travel effect (LTTE) component of the ETV residual signal.
Using the same form as \citet{Rapp}, the timings can be expressed by

\begin{equation}
	ETV_\text{LTTE} = A_{LTTE} \left[ \left( 1 - e_3^2 \right)^{1/2} \sin E_3(t) \cos \omega_3 + \left( \cos E_3(t) -e_3 \right) \sin \omega_3  \right]
\end{equation}
where
\begin{align}
	E_3(t) &= M_3(t) + e_3 \sin E_3(t) \\
	M_3(t) &= \left( t - t_0 \right) \frac{2 \pi}{P_3} \\
	A_{LTTE} &= \frac{G^{1/3}}{c (2 \pi )^{2/3}} \left[ \frac{m_3}{m_{123}^{2/3}} \sin i_3 \right] P_3^{2/3} 
\end{align}
and $t_0$ is a time offset,  $m_3$ is the mass of the third body, $m_{123}$ is the mass of the entire system, and $P_3$, $i_3$, $e_3$, $\omega_3$, $E_3(t)$, and $M_3(t)$ are the period, inclination, eccentricity, argument of periastron, eccentric anomaly, and mean anomaly of the third body orbit, respectively.

This expression was then used to fit all ETVs flagged as potential third body signals.
The period was first estimated using Lomb-Scargle periodogram and used as input into a series of Levenberg-Marquardt fits, each using a different starting guess for eccentricity.
The fit with the lowest chi-squared was then kept and the errors estimated from the covariance matrix.
If the final fit had an eccentricity consistent with 0, then $e_3$ was set to 0, $\omega_3$ to $\pi/2$, and the fitting was redone with circular constraints to get appropriate error estimates on the remaining parameters.

\begin{figure}
\plotone{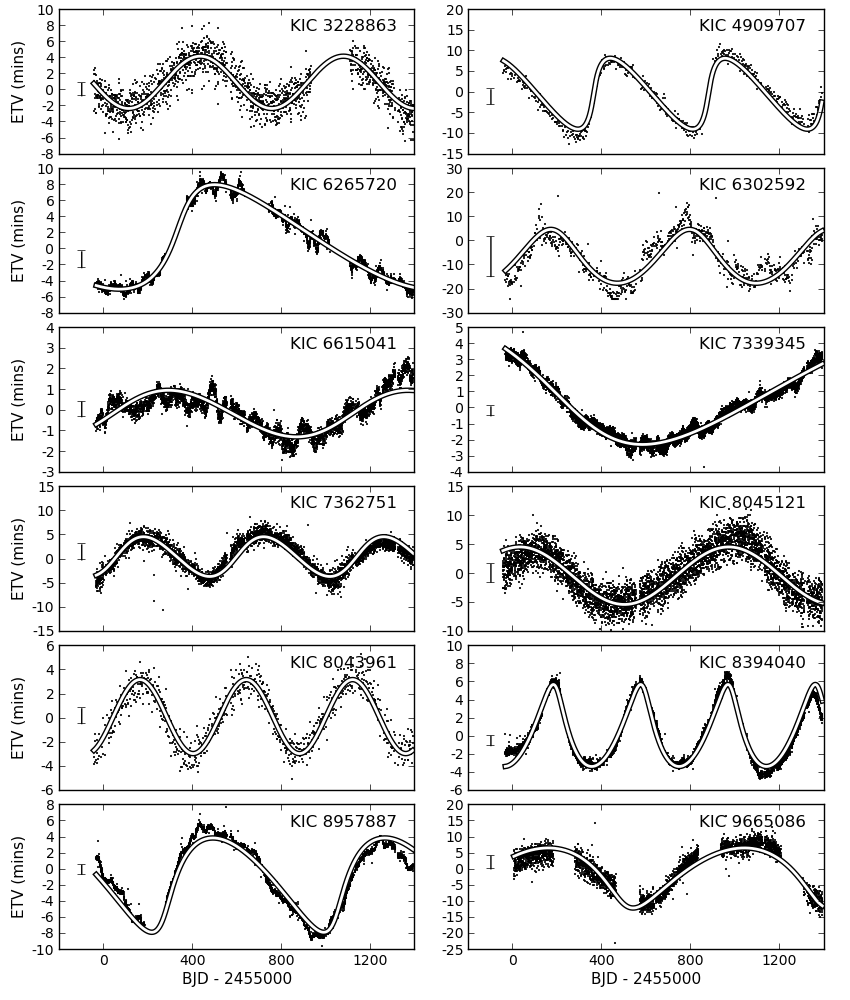}
\caption{Gallery of select ETV signals found in close binaries with LTTE fits.  
These are KIC 3228863, 4909707, 6265720, 6302592, 6615041, 7339345, 7362751, 8045121, 8043961, 8394040, 8957887, and 9665086.  
Typical errors for ETV measurements are shown to the left of the data.}
\label{etv_LTTE}
\end{figure}

This gives values and estimated errors for $P_3$, $e_3$, and $A_{LTTE}$ (Table \ref{tableresults}).
A sample of some of these ETV signals and their respective fits can be seen in Fig.~\ref{etv_LTTE}, with fits for all candidate third body signals available in the online version of the \emph{Kepler} EB Catalog.  
We can only provide estimate periods for the sample of ETV signals with less than one full cycle in the data.  Even these periods can be significantly biased based on the section of the cycle that is in the observed baseline and should be treated with reservation.  These very long period cases are provided separately at the end of Table \ref{tableresults}

\subsection{Physical Delay}

\begin{figure}
\plotone{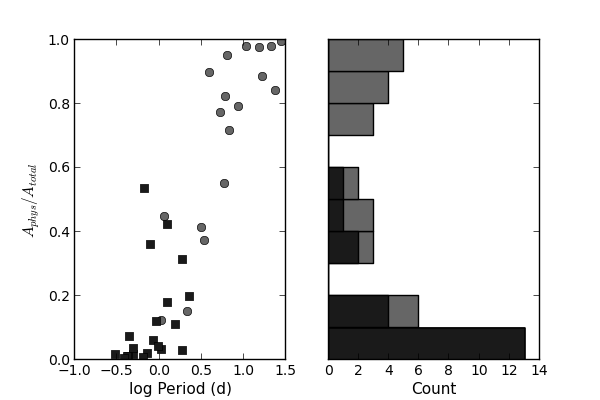}
\caption{Distribution of the contribution to the total ETV amplitude due to physical effects \citep{Rapp} .  
The left plot shows this distribution versus the log of the period of the inner-binary, with the systems that overlap with this paper as dark squares, and those that do not as lighter circles.
The right panel shows a histogram of these contributions, with the overlapped systems highlighted with the darker shade.}
\label{phys_hist}
\end{figure}

\citet{Rapp} included physical delays in their models of 39 \emph{Kepler} binaries with possible third-body ETVs, sometimes contributing largely to the overall model.
This dynamical effect occurs when the presence of a third body changes the period of the inner binary.
Fig.~\ref{phys_hist} shows the distribution in their targets and the overlapping targets of the ratio of the amplitude of the physical delay compared to the total amplitude in the ETV signal.    
21 of their targets overlap with ours, but due to the short-period inner-binary, the physical delay rarely contributes significantly to these model. 
From their results, it seems that the LTTE effect dominates over the physical delay for binaries with periods less than 3 days, which covers the vast majority of our targets. 

\subsection{Objects with Tertiary Eclipses}

For some of these binaries with LTTE signals, tertiary eclipses have also been found that confirm the presence and third body period, and significantly constrain the inclination of the third body.
Any binary which was identified to have a possible third body due to its ETV signal and also has a detected tertiary eclipse is noted as such in Table \ref{tableresults}.

\begin{figure}
\plotone{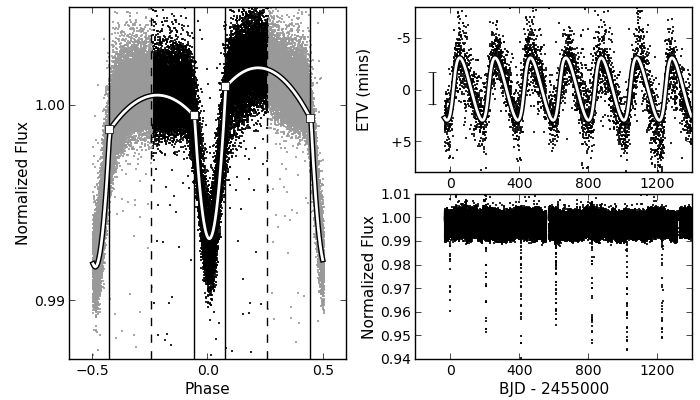}
\caption{A triple eclipsing star KIC 2856960. Left: the detrended light curve phased at the inner
period of 0.26-d. The white line is the polyfit function, and white rectangles are the knots. Dashed
lines delimit the phase space of the primary and secondary eclipse; these are used separately to
obtain primary and secondary ETVs. Upper right: the measured ETVs (black points) and the best
light-time travel fit (white line), yielding the outer period of 205.5 days. Lower right: the detrended light
curve, with the tertiary eclipses clearly visible.}
\label{2856960}
\end{figure}

KIC 2856960, for instance, has an inner-binary period of 0.259 days with an ETV signal resulting in a LTTE fit with a period of $205.5 \pm 0.1$ days.
This period is consistent with the previously determined period for the tertiary events of 204.25 days (Fig.~\ref{2856960}).
This is also consistent with the LTTE period of $205 \pm 2$ days reported by \citet{Lee}, and the tertiary event period of $204.2$ days by \citet{Armstrong}.

\begin{figure}
\plotone{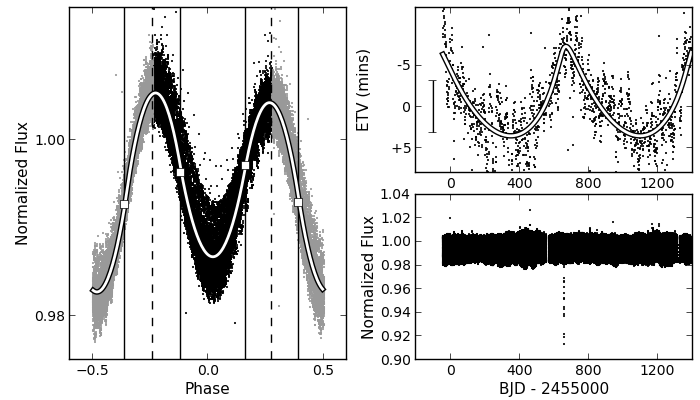}
\caption{KIC 2835289 is an ellipsoidal variable with a period of 0.857 days.  We can see one tertiary eclipse in the light curve and the ETV signal can put an additional constraint on the expected period of a potential third-body.}
\label{2835289}
\end{figure}

In the case of KIC 2835289 (Fig.~\ref{2835289}), we have only observed one potential tertiary event in Q9.  
Without at least three consecutive events, we cannot rigorously confirm that the eclipse is a third-body as opposed to a blended eclipsing binary.
However, the eclipse seems to show the eclipse of both stars in the inner-binary and ETV signal shows a possible long-term third body orbit suggesting a period of approximately 800 days.
If this proves to be a true third body, then \emph{Kepler} just missed an event before the beginning of the mission and may have observed another event in Q17, which has yet to be processed.

KIC 6543674 also shows a single tertiary eclipse in Q2.
A second tertiary eclipse was missed during a break in the \emph{Kepler} data, but we were able to observe an additional tertiary event from the ground, giving a third body period of $\sim 1100$ days \citep{Thackeray-Lacko}.
In this case, we do not have a full orbit of the ETV signal and the LTTE model period is quite uncertain.

\subsection{Objects with Depth Variations}

\begin{figure}
\plotone{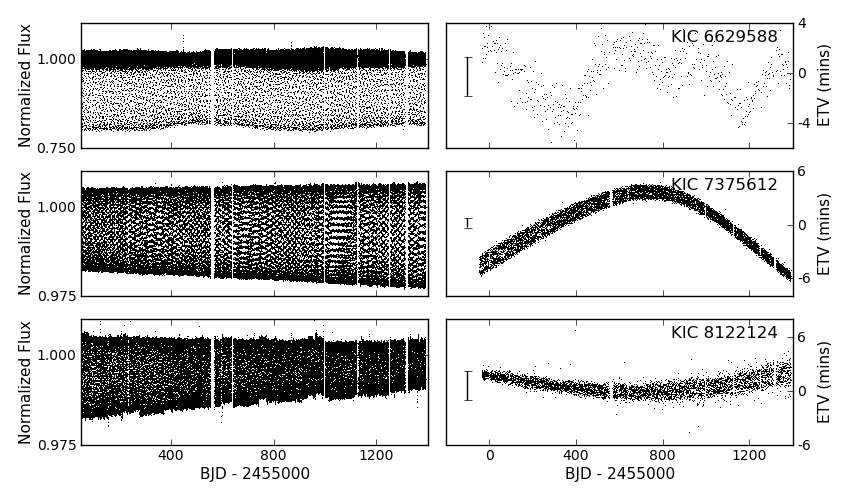}
\caption{KIC \DVKICs{} all show ETVs that suggest a possible third body and eclipse depth variations.  In some of these cases the template polyfit causes systematics in the timings, and could be improved by creating new templates for each quarter.}
\label{DV}
\end{figure}

\numberDV{} binaries that show third-body ETV signals (KIC \DVKICs{}) also show constant changes in their eclipse depths (Fig.~\ref{DV}), which could either be caused by a change in inclination or apsidal motion perhaps induced by the third body.
We plan to follow these up later with full photodynamical models.

\subsection{Potential Fourth Body Signals}

\begin{figure}
\plotone{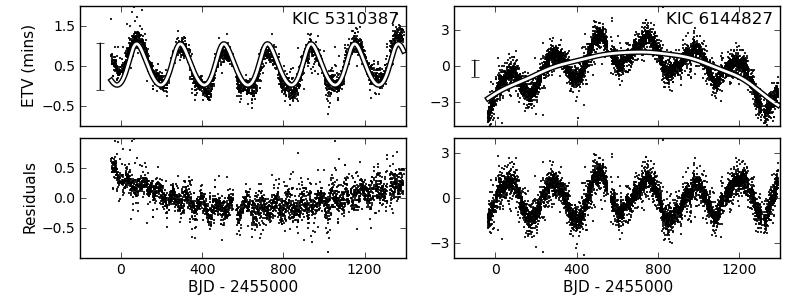}
\caption{KIC 5310387 and 6144827 are among several ETV signals with residuals that suggest another parabolic or LTTE signal, possibly indicating the presence of a fourth body.}
\label{quad}
\end{figure}

It is also possible that some of these ETVs could be composed of multiple signals.
KIC 5310387, 6144827, 8145477, 11612091, and 11825204, for example, may have both an LTTE and quadratic component or two LTTE signals as is shown in the residuals in Fig.~\ref{quad}.
In general, the stronger signal is fitted and noted.

\section{Discussion}

\begin{figure}
\plotone{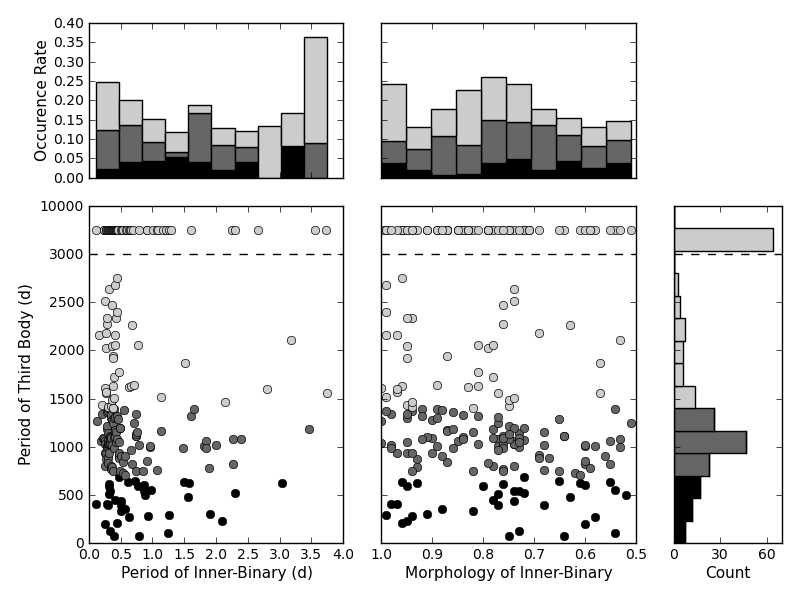}
\caption{Distribution of period of potential third body companions verses the inner-binary period (left) and morphology (right).
Third body periods greater than 3000 days are all placed in the final bin despite their modeled periods.
The different colors represent the three different samples of binaries represented in Table \ref{tableresults}, determined by the period of the potential third body.
The top histograms show the occurrence rate of candidate third bodies for each bin in period or morphology, and the histogram on the right shows the number of third body candidates at each period.
}
\label{ptrip}
\end{figure}

In this study we find a third body rate of \rateTriple{} in our sample of close binaries, nearly all of which have inner binary periods shorter than 3 days (Fig.~\ref{ptrip}).
This is much lower than the third body rate of 96\% found by the previous studies mentioned.
However, our identification of tertiary companions is certainly a lower limit for several reasons.

First, our ability to detect a third body is very sensitive to both inclination and mass of the third body,
such that low-mass tertiaries and/or tertiaries whose orbital planes are highly inclined relative to the inner binary orbital plane do not present detectable LTTE effects.
Of our total sample of \numberShortEBs{} binaries, \numberTertiaryEclipse{} (\rateTertiaryEclipse{}) show an LTTE orbit and visible tertiary eclipses.
\numberTMsectionAB{} (\rateTMsectionAB{}) have LTTE orbits with periods shorter than the span of our photometric data but do not show tertiary eclipses, suggesting that the eclipses fell in a gap in the data or the orbits are not well enough aligned to show eclipses. Thus there is evidence from these examples that 
in a few percent of cases we are indeed missing true third bodies because of inclination non-alignments.
\numberTMsectionC{} (\rateTMsectionC{}) have LTTE orbits with periods longer than the photometric baseline. In these cases we do not have well constrained periods and our chances of detecting a tertiary eclipse are slim.

A second reason that our determination of the third-body occurrence is likely a lower limit is that 
the very close binaries that comprise our sample here generally present more noise in the ETV signal, which could easily bury a weak LTTE signal. 
We have employed a method that minimizes false positives due to spurious ETV signals, and thus necessarily have eliminated some potentially true LTTE signals.

Third, and perhaps most important, the limited timespan of the currently available \emph{Kepler} data ($\sim$1400 days) significantly restricts 
us to detect third bodies with orbital periods comparable to or shorter than 1400 days. Relative to the full span of tertiary separations found in 
previous works \citep{Tok97,Tok06,slowpokes,law}, with separations as large as $\sim$1 pc, we are at present sampling only
the relatively closest tertiary companions.
Indeed, \citet{Tok06} found among tight binaries that the rate of third bodies with orbital periods less than $\sim$3 years (comparable to our limit based on the duration of the available \emph{Kepler} data) is 15\% $\pm$ 3\%.
Thus our finding of a third-body occurrence rate with a period less than 1400 days of \rateTMsectionAB{} is compatible with the expected rate, though it appears we are likely still missing a fraction of some systems for the reasons already mentioned.

The distribution of periods of potential third body orbits is also shown in Fig.~\ref{ptrip}.
We can clearly see a falloff in detection past the current length of the \emph{Kepler} mission of $\sim$ 1400 days, as expected. 
However, for third-body periods shorter than $\sim$ 1400 days, for which our detectability is relatively good, 
the occurrence rate does appear to increase toward longer third-body periods,
consistent with the period distribution of third bodies among tight binaries
found by \citet{Tok06}.
Furthermore, we find that the triples on the widest orbits are found around the shortest period binaries, which is consistent with models that tighten the inner binary orbit through the presence, and gradual widening, of a companion.

\section{Summary and Conclusions}

We presented our technique for computing precise eclipse timings for \numberShortEBs{} close eclipsing binaries in the \emph{Kepler} Eclipsing Binary Catalog. 
These precise eclipse timings are complemented by the eclipse timings to be reported by \citet{KepEBetv1} for longer period, detached EBs. 
For the EBs whose timings are reported here, our method has been developed specifically to deal with the challenge of constantly changing light levels arising from
spots and other phenomena that distort the light curves and could cause spurious eclipse timing variation (ETV) signals.

EBs with ETV signals suggesting the possible presence of a third body have been identified and have been fit with a LTTE orbit model in order to determine the likely parameters of the third bodies.
In the current sample of \numberShortEBs{} close EBs, we have identified \numberTM{} that likely have tertiary companions.
The parameters of these fits are also available online and are updated as new data become available.

Our measured occurrence rate of \rateTMsectionAB{} of close binaries with tertiary companions with periods up to $\sim 1400$ days (limited by the current timespan of \emph{Kepler} data), appears to be broadly consistent with the
expectation that $15 \pm 3$\% of close binaries will have tertiaries of such periods \citep{Tok06}. 
Indeed, we already find in our data that the periods of third bodies rise among the tightest binaries, consistent with previous
work that has found a very high rate of third bodies in very wide orbits around the tightest binaries, presumably the result of dynamical tightening of inner binaries through widening of the tertiary.

Eclipse timings for all EBs are updated in real-time and are freely available as a community resource at \texttt{http://keplerEBs.villanova.edu}.

\section*{Acknowledgements}

We thank Nathan De Lee and Phil Cargile for helpful discussions and Darin Ragozzine, Eric Ford, and Joshua Pepper for their feedback.

This project is supported through the \emph{Kepler} Participating Scientist Award NSR303065.
KEC and KGS acknowledge support from NASA ADAP grant NNX12AE22G.
JAO and WFW gratefully acknowledge support from the NSF via grant AST-1109928, and also NASA via the \emph{Kepler} PSP grant NNX12AD23G. 

\emph{Kepler} was selected as the 10th mission of the Discovery Program. Funding for this mission is provided by NASA, Science Mission Directorate.

\begin{deluxetable}{rrrrr}
\tablecolumns{5}
\tablewidth{0pc}
\tablecaption{Eclipse Timing Variations  \label{tableetvs}}
\tablehead{
	\colhead{KIC} & \colhead{$BJD_{ecl}$} & \colhead{$ETV$ (s)} & \colhead{$\sigma ({ETV})$ (s)} & \colhead{Eclipse} 
}
\startdata
1433410 & 56107.275 & -6.307 & 39.658 & primary \\ 
1433410 & 56107.275 & -4.493 & 20.390 & entire \\ 
1433410 & 56107.416 & -47.606 & 40.954 & secondary \\ 
1433410 & 56107.558 & -10.541 & 18.835 & entire \\ 
1433410 & 56107.558 & 21.341 & 30.499 & primary \\ 
1433980 & 55740.148 & -443.232 & 143.078 & secondary \\ 
1433980 & 55740.944 & -283.392 & 70.502 & primary \\ 
1433980 & 55740.944 & -170.899 & 69.552 & entire \\ 
1433980 & 55741.741 & -161.482 & 90.634 & secondary \\ 
1433980 & 55742.537 & -248.659 & 61.776 & primary \\ 
\enddata
\tablecomments{The full electronic table is available in the online version.}
\end{deluxetable}

\begin{deluxetable}{rrr}
\tablecolumns{3}
\tablewidth{0pc}
\tablecaption{ETVs with Parabolic Signals}  \label{tableMT}
\tablehead{
	\colhead{KIC} & \colhead{KIC} & \colhead{KIC} 
}
\startdata
2305372 & 3104113 & 3765708 \\ 
4074532 & 4851217 & 4853067 \\ 
5020034 & 5471619 & 5770860 \\ 
5792093 & 6044064 & 6044543 \\ 
6066379 & 6213131 & 6314173 \\ 
6464285 & 6677225 & 7696778 \\ 
7938468 & 7938870 & 8758161 \\ 
9087918 & 9402652 & 9840412 \\ 
9934052 & 10030943 & 10292413 \\ 
10736223 & 11097678 & 11144556 \\ 
11924311 & \nodata & \nodata \\ 
\enddata
\end{deluxetable}

\begin{deluxetable}{rrrrrrr}
\tablecolumns{7}
\tablewidth{0pc}
\tablecaption{ETVs crossmatched with \citet{Rapp}  \label{tablerapp}}
\tablehead{
	\colhead{KIC} & \colhead{$P_{3,R}$ (d)} & \colhead{$P_{3}$ (d)} & \colhead{$e_{3,R}$} & \colhead{$e_{3}$} & \colhead{$A_{LTTE,R} (s)$} & \colhead{$A_{LTTE} (s)$} 
}
\startdata
3228863 & 668.4 & $644.1 \pm 15.7$ & 0.08\{0.06,0.12\} & $0.000 \pm 0.003$ & 189\{187,194\} & $195 \pm 3$ \\
4647652 & 753.5 & $755.2 \pm 44.3$ & 0.35\{0.10,0.44\} & $0.244 \pm 0.003$ & 228\{183,274\} & $239 \pm 9$ \\
4909707 & 505.3 & $516.1 \pm 16.1$ & 0.54\{0.31,0.66\} & $0.686 \pm 0.006$ & 493\{378,627\} & $707 \pm 14$ \\
5128972 & 447.8 & $438.7 \pm 1.9$ & 0.33\{0.25,0.41\} & $0.323 \pm 0.002$ & 259\{244,271\} & $256 \pm 1$ \\
5264818 & 296.3 & $299.7 \pm 107.5$ & 0.37\{0.13,0.53\} & $0.421 \pm 0.306$ & 145\{107,196\} & $178 \pm 42$ \\
5310387 & 214.2 & $214.3 \pm 0.3$ & 0.53\{0.34,0.61\} & $0.250 \pm 0.004$ & 31\{ 27, 37\} & $31 \pm 1$ \\
5376552 & 334.5 & $331.1 \pm 0.8$ & 0.40\{0.35,0.45\} & $0.000 \pm 0.002$ & 94\{ 91, 98\} & $87 \pm 1$ \\
6370665 & 285.9 & $283.2 \pm 20.9$ & 0.22\{0.07,0.33\} & $0.136 \pm 0.085$ & 67\{ 61, 74\} & $66 \pm 3$ \\
6531485 & 48.3 & \nodata & 0.44\{0.33,0.63\} & \nodata & 72\{ 31,109\} & \nodata \\
7690843 & 74.3 & $74.1 \pm 0.1$ & 0.25\{0.08,0.42\} & $0.233 \pm 0.021$ & 71\{ 51, 91\} & $81 \pm 1$ \\
8043961 & 476.7 & $478.0 \pm 10.4$ & 0.25\{0.14,0.33\} & $0.000 \pm 0.005$ & 194\{179,213\} & $184 \pm 2$ \\
8192840 & 803.9 & $1045.9 \pm 185.0$ & 0.63\{0.52,0.70\} & $0.616 \pm 0.002$ & 208\{187,223\} & $260 \pm 30$ \\
8386865 & 293 & $293.9 \pm 2.8$ & 0.38\{0.27,0.48\} & $0.493 \pm 0.013$ & 171\{156,210\} & $197 \pm 1$ \\
8394040 & 394.8 & $392.6 \pm 0.8$ & 0.61\{0.50,0.67\} & $0.467 \pm 0.001$ & 369\{345,391\} & $278 \pm 1$ \\
8904448 & 548.1 & $538.8 \pm 59.9$ & 0.59\{0.50,0.66\} & $0.577 \pm 0.016$ & 171\{158,192\} & $166 \pm 12$ \\
9451096 & 106.7 & $106.8 \pm 0.1$ & 0.24\{0.10,0.36\} & $0.091 \pm 0.033$ & 90\{ 59,144\} & $93 \pm 1$ \\
9722737 & 443.9 & $451.3 \pm 3.7$ & 0.22\{0.16,0.27\} & $0.152 \pm 0.003$ & 230\{225,236\} & $225 \pm 1$ \\
9912977 & 753.7 & $780.4 \pm 95.2$ & 0.31\{0.16,0.39\} & $0.504 \pm 0.008$ & 105\{ 94,117\} & $96 \pm 7$ \\
10226388 & 934.9 & $965.3 \pm 183.8$ & 0.32\{0.24,0.39\} & $0.041 \pm 0.007$ & 465\{434,493\} & $457 \pm 58$ \\
10991989 & 554.2 & $554.8 \pm 64.1$ & 0.30\{0.21,0.37\} & $0.000 \pm 0.018$ & 256\{239,274\} & $232 \pm 17$ \\
11042923 & 839 & $984.4 \pm 63.9$ & 0.17\{0.09,0.25\} & $0.258 \pm 0.002$ & 223\{213,230\} & $276 \pm 11$ \\
\enddata
\tablecomments{$P_{3,R}$, $e_{3,R}$, and $A_{LTTE,R}$ are the period, eccentricity, and amplitude as reported by \citet{Rapp}}
\end{deluxetable}

\begin{deluxetable}{rrrrrr}
\tablecolumns{6}
\tablewidth{0pc}
\tablecaption{ETVs with Potential Third-Body Signals  \label{tableresults}}
\tablehead{
	\colhead{KIC} & \colhead{$\text{morph}_{bin}$} & \colhead{$P_{bin}$ (d)}   & \colhead{$P_{3}$ (d)} & \colhead{$e_{3}$} & \colhead{$A_{LTTE} (s)$} 
}
\startdata
2856960\tablenotemark{5} & 0.60 & 0.259 & $204.5 \pm 0.1$ & $0.447 \pm 0.001$ & $202 \pm 1$ \\
3228863\tablenotemark{1} & 0.65 & 0.731 & $644.1 \pm 15.7$ & $0.000 \pm 0.003$ & $195 \pm 3$ \\
3245776 & 0.96 & 1.492 & $636.3 \pm 70.6$ & $0.587 \pm 0.021$ & $136 \pm 10$ \\
3641446 & 0.95 & 2.100 & $228.6 \pm 1.0$ & $0.000 \pm 0.010$ & $85 \pm 1$ \\
4037163 & 0.58 & 0.635 & $267.0 \pm 8.1$ & $0.349 \pm 0.009$ & $77 \pm 1$ \\
4909707\tablenotemark{1} & 0.72 & 2.302 & $516.1 \pm 16.1$ & $0.686 \pm 0.006$ & $707 \pm 14$ \\
5128972\tablenotemark{1} & 0.74 & 0.505 & $438.7 \pm 1.9$ & $0.323 \pm 0.002$ & $256 \pm 1$ \\
5264818\tablenotemark{1} & 0.91 & 1.905 & $299.7 \pm 107.5$ & $0.421 \pm 0.306$ & $178 \pm 42$ \\
5310387\tablenotemark{1}\tablenotemark{4} & 0.96 & 0.442 & $214.3 \pm 0.3$ & $0.250 \pm 0.004$ & $31 \pm 1$ \\
5376552\tablenotemark{1} & 0.82 & 0.504 & $331.1 \pm 0.8$ & $0.000 \pm 0.002$ & $87 \pm 1$ \\
5459373 & 0.97 & 0.287 & $411.5 \pm 1.2$ & $0.372 \pm 0.002$ & $228 \pm 1$ \\
5560831 & 0.60 & 0.868 & $609.0 \pm 149.2$ & $0.093 \pm 0.010$ & $58 \pm 9$ \\
6302592 & 0.93 & 1.578 & $623.6 \pm 42.9$ & $0.211 \pm 0.009$ & $671 \pm 30$ \\
6370665\tablenotemark{1} & 0.94 & 0.932 & $283.2 \pm 20.9$ & $0.136 \pm 0.085$ & $66 \pm 3$ \\
7362751 & 0.73 & 0.338 & $540.3 \pm 3.4$ & $0.162 \pm 0.001$ & $250 \pm 1$ \\
7657914 & 0.72 & 0.475 & $689.9 \pm 295.1$ & $0.405 \pm 0.025$ & $30 \pm 8$ \\
7685689 & 0.77 & 0.325 & $507.3 \pm 6.2$ & $0.176 \pm 0.002$ & $183 \pm 1$ \\
7690843\tablenotemark{1} & 0.64 & 0.786 & $74.1 \pm 0.1$ & $0.233 \pm 0.021$ & $81 \pm 1$ \\
8043961\tablenotemark{1} & 0.63 & 1.559 & $478.0 \pm 10.4$ & $0.000 \pm 0.005$ & $184 \pm 2$ \\
8145477 & 0.88 & 0.566 & $353.7 \pm 46.7$ & $0.418 \pm 0.007$ & $136 \pm 12$ \\
8190491 & 0.95 & 0.778 & $594.7 \pm 11.7$ & $0.000 \pm 0.003$ & $130 \pm 1$ \\
8211618 & 0.73 & 0.337 & $127.3 \pm 66.7$ & $0.319 \pm 0.137$ & $31 \pm 10$ \\
8330092 & 0.80 & 0.322 & $595.5 \pm 5.4$ & $0.201 \pm 0.001$ & $127 \pm 1$ \\
8386865\tablenotemark{1} & 0.99 & 1.258 & $293.9 \pm 2.8$ & $0.493 \pm 0.013$ & $197 \pm 1$ \\
8394040\tablenotemark{1} & 0.77 & 0.302 & $392.6 \pm 0.8$ & $0.467 \pm 0.001$ & $278 \pm 1$ \\
8904448\tablenotemark{1} & 0.74 & 0.866 & $538.8 \pm 59.9$ & $0.577 \pm 0.016$ & $166 \pm 12$ \\
9075704 & 0.68 & 0.513 & $396.3 \pm 7.5$ & $0.101 \pm 0.003$ & $138 \pm 1$ \\
9451096\tablenotemark{1} & 0.54 & 1.250 & $106.8 \pm 0.1$ & $0.091 \pm 0.033$ & $93 \pm 1$ \\
9706078 & 0.55 & 0.614 & $639.2 \pm 27.6$ & $0.550 \pm 0.004$ & $237 \pm 6$ \\
9722737\tablenotemark{1} & 0.78 & 0.419 & $451.3 \pm 3.7$ & $0.152 \pm 0.003$ & $225 \pm 1$ \\
9994475 & 0.76 & 0.318 & $610.9 \pm 6.8$ & $0.375 \pm 0.001$ & $196 \pm 1$ \\
10014830\tablenotemark{4} & 0.61 & 3.031 & $625.0 \pm 44.6$ & $0.487 \pm 0.012$ & $156 \pm 7$ \\
10855535\tablenotemark{4} & 0.98 & 0.113 & $411.8 \pm 0.5$ & $0.095 \pm 0.001$ & $142 \pm 1$ \\
10991989\tablenotemark{1} & 0.54 & 0.974 & $554.8 \pm 64.1$ & $0.000 \pm 0.018$ & $232 \pm 17$ \\
11247386 & 0.75 & 0.394 & $71.2 \pm 0.1$ & $0.217 \pm 0.011$ & $38 \pm 1$ \\
\tableline 
2302092 & 0.89 & 0.295 & $1010.7 \pm 48.7$ & $0.468 \pm 0.002$ & $435 \pm 13$ \\
2450566 & 0.98 & 1.845 & $983.7 \pm 472.8$ & $0.308 \pm 0.016$ & $431 \pm 138$ \\
2835289\tablenotemark{5} & 0.94 & 0.858 & $747.4 \pm 23.7$ & $0.643 \pm 0.003$ & $338 \pm 7$ \\
3839964 & 0.78 & 0.256 & $798.2 \pm 167.5$ & $0.530 \pm 0.016$ & $51 \pm 7$ \\
4069063\tablenotemark{4} & 0.56 & 0.504 & $906.3 \pm 26.2$ & $0.516 \pm 0.002$ & $430 \pm 8$ \\
4138301 & 0.90 & 0.253 & $934.1 \pm 65.3$ & $0.272 \pm 0.002$ & $329 \pm 15$ \\
4244929 & 0.91 & 0.341 & $1103.1 \pm 38.5$ & $0.619 \pm 0.001$ & $228 \pm 5$ \\
4451148 & 0.82 & 0.736 & $746.0 \pm 52.1$ & $0.293 \pm 0.004$ & $322 \pm 14$ \\
4547308 & 0.88 & 0.577 & $908.6 \pm 75.3$ & $0.000 \pm 0.003$ & $154 \pm 8$ \\
4647652\tablenotemark{1} & 0.68 & 1.065 & $755.2 \pm 44.3$ & $0.244 \pm 0.003$ & $239 \pm 9$ \\
4670267 & 0.60 & 2.006 & $1017.4 \pm 429.9$ & $0.751 \pm 0.011$ & $82 \pm 23$ \\
4681152 & 0.55 & 1.836 & $1063.2 \pm 510.0$ & $0.514 \pm 0.018$ & $88 \pm 28$ \\
4762887 & 0.95 & 0.737 & $1340.6 \pm 849.3$ & $0.000 \pm 0.005$ & $52 \pm 22$ \\
4859432 & 0.76 & 0.385 & $749.6 \pm 12.9$ & $0.591 \pm 0.001$ & $154 \pm 1$ \\
4937217 & 0.82 & 0.429 & $1152.6 \pm 490.9$ & $0.368 \pm 0.013$ & $26 \pm 7$ \\
4945857 & 0.74 & 0.335 & $1026.6 \pm 66.7$ & $0.000 \pm 0.002$ & $578 \pm 25$ \\
5269407 & 0.53 & 0.959 & $1003.2 \pm 114.3$ & $0.000 \pm 0.003$ & $217 \pm 16$ \\
5478466 & 0.97 & 0.483 & $934.5 \pm 98.2$ & $0.754 \pm 0.002$ & $262 \pm 18$ \\
5611561 & 0.74 & 0.259 & $1033.8 \pm 265.3$ & $0.000 \pm 0.005$ & $103 \pm 17$ \\
5790912 & 0.77 & 0.383 & $1245.9 \pm 798.3$ & $0.677 \pm 0.009$ & $64 \pm 27$ \\
5791886 & 0.76 & 0.325 & $1032.2 \pm 63.0$ & $0.937 \pm 0.007$ & $69 \pm 2$ \\
5975712 & 0.87 & 1.136 & $1164.7 \pm 964.3$ & $0.000 \pm 0.013$ & $424 \pm 234$ \\
6050116 & 0.77 & 0.240 & $1078.4 \pm 399.2$ & $0.000 \pm 0.009$ & $53 \pm 13$ \\
6118779 & 0.90 & 0.364 & $1281.4 \pm 49.4$ & $0.972 \pm 0.003$ & $235 \pm 6$ \\
6281103 & 0.98 & 0.363 & $1018.6 \pm 40.5$ & $0.153 \pm 0.001$ & $132 \pm 3$ \\
6469946 & 0.51 & 0.716 & $1246.5 \pm 518.1$ & $0.978 \pm 0.005$ & $1803 \pm 499$ \\
6516874 & 0.60 & 0.916 & $857.9 \pm 376.5$ & $0.000 \pm 0.016$ & $194 \pm 56$ \\
6543674\tablenotemark{3}\tablenotemark{5} & 0.53 & 2.391 & $1085.3 \pm 224.9$ & $0.593 \pm 0.008$ & $263 \pm 36$ \\
6615041 & 0.75 & 0.340 & $1077.5 \pm 42.4$ & $0.107 \pm 0.001$ & $68 \pm 1$ \\
6629588\tablenotemark{4} & 0.55 & 2.264 & $818.7 \pm 93.5$ & $0.424 \pm 0.006$ & $128 \pm 9$ \\
6671698 & 0.73 & 0.472 & $1048.0 \pm 63.2$ & $0.105 \pm 0.002$ & $193 \pm 7$ \\
6766325 & 0.92 & 0.440 & $1316.6 \pm 921.7$ & $0.592 \pm 0.007$ & $138 \pm 64$ \\
7035139 & 0.79 & 0.310 & $831.5 \pm 38.1$ & $0.479 \pm 0.002$ & $62 \pm 1$ \\
7119757 & 0.64 & 0.743 & $1109.4 \pm 398.5$ & $0.666 \pm 0.006$ & $427 \pm 102$ \\
7272739 & 0.75 & 0.281 & $1220.8 \pm 354.0$ & $0.554 \pm 0.006$ & $83 \pm 16$ \\
7385478 & 0.54 & 1.655 & $1389.3 \pm 795.2$ & $0.245 \pm 0.007$ & $243 \pm 93$ \\
7518816 & 0.65 & 0.467 & $1283.9 \pm 618.5$ & $0.481 \pm 0.001$ & $73 \pm 23$ \\
7877062 & 0.81 & 0.304 & $1024.4 \pm 31.0$ & $0.169 \pm 0.001$ & $93 \pm 1$ \\
8045121 & 0.99 & 0.263 & $938.6 \pm 25.8$ & $0.000 \pm 0.001$ & $298 \pm 5$ \\
8192840\tablenotemark{1} & 0.95 & 0.434 & $1045.9 \pm 185.0$ & $0.616 \pm 0.002$ & $260 \pm 30$ \\
8242493 & 0.73 & 0.283 & $993.5 \pm 101.0$ & $0.000 \pm 0.003$ & $65 \pm 4$ \\
8563964 & 1.00 & 0.338 & $1035.2 \pm 49.7$ & $0.000 \pm 0.001$ & $195 \pm 6$ \\
8690104 & 0.77 & 0.409 & $1304.2 \pm 408.9$ & $0.811 \pm 0.009$ & $77 \pm 16$ \\
8739802 & 0.93 & 0.275 & $869.9 \pm 19.9$ & $0.302 \pm 0.001$ & $125 \pm 1$ \\
8957887 & 0.76 & 0.347 & $773.1 \pm 12.5$ & $0.561 \pm 0.001$ & $401 \pm 4$ \\
8982514 & 0.84 & 0.414 & $1106.3 \pm 61.1$ & $0.000 \pm 0.001$ & $58 \pm 2$ \\
9091810 & 0.69 & 0.480 & $888.9 \pm 55.4$ & $0.000 \pm 0.003$ & $75 \pm 3$ \\
9101279 & 0.58 & 1.811 & $1010.9 \pm 1000.2$ & $0.212 \pm 0.021$ & $78 \pm 51$ \\
9272276 & 0.78 & 0.281 & $1187.3 \pm 133.6$ & $0.347 \pm 0.001$ & $458 \pm 34$ \\
9283826 & 0.84 & 0.357 & $1334.5 \pm 381.1$ & $0.000 \pm 0.002$ & $169 \pm 32$ \\
9353234 & 0.86 & 1.487 & $983.9 \pm 352.2$ & $0.114 \pm 0.010$ & $132 \pm 31$ \\
9412114 & 0.85 & 0.250 & $1060.9 \pm 1527.5$ & $0.330 \pm 0.005$ & $450 \pm 432$ \\
9532219 & 0.74 & 0.198 & $1062.1 \pm 76.3$ & $0.372 \pm 0.002$ & $70 \pm 3$ \\
9592145 & 0.65 & 0.489 & $748.7 \pm 1091.9$ & $0.396 \pm 0.086$ & $14 \pm 14$ \\
9612468 & 1.00 & 0.133 & $1264.2 \pm 233.2$ & $0.340 \pm 0.001$ & $118 \pm 14$ \\
9665086\tablenotemark{5} & 0.67 & 0.297 & $882.3 \pm 36.9$ & $0.447 \pm 0.002$ & $576 \pm 16$ \\
9821923 & 0.95 & 0.350 & $1295.2 \pm 354.8$ & $0.615 \pm 0.004$ & $216 \pm 39$ \\
9838047 & 0.84 & 0.436 & $1082.2 \pm 40.7$ & $0.353 \pm 0.001$ & $490 \pm 12$ \\
9912977\tablenotemark{1} & 0.59 & 1.888 & $780.4 \pm 95.2$ & $0.504 \pm 0.008$ & $96 \pm 7$ \\
10226388\tablenotemark{1} & 0.77 & 0.661 & $965.3 \pm 183.8$ & $0.041 \pm 0.007$ & $457 \pm 58$ \\
10275197 & 0.78 & 0.391 & $1093.6 \pm 28.5$ & $0.516 \pm 0.001$ & $278 \pm 4$ \\
10322582 & 0.86 & 0.291 & $1362.9 \pm 349.1$ & $0.697 \pm 0.003$ & $354 \pm 60$ \\
10383620 & 0.64 & 0.735 & $1111.3 \pm 253.4$ & $0.000 \pm 0.007$ & $457 \pm 69$ \\
10388897 & 0.99 & 0.344 & $1367.6 \pm 1366.9$ & $0.586 \pm 0.005$ & $301 \pm 200$ \\
10724533 & 0.75 & 0.745 & $1131.4 \pm 197.7$ & $0.265 \pm 0.003$ & $73 \pm 8$ \\
10727655 & 0.73 & 0.353 & $1087.9 \pm 67.6$ & $0.000 \pm 0.001$ & $309 \pm 12$ \\
10848807 & 0.74 & 0.346 & $799.1 \pm 135.4$ & $0.430 \pm 0.011$ & $31 \pm 3$ \\
10905804 & 0.68 & 0.751 & $1154.1 \pm 471.0$ & $0.484 \pm 0.022$ & $49 \pm 13$ \\
10916675 & 0.87 & 0.419 & $1170.9 \pm 244.5$ & $0.609 \pm 0.007$ & $29 \pm 4$ \\
10934755 & 0.68 & 0.786 & $1021.3 \pm 112.6$ & $0.302 \pm 0.005$ & $84 \pm 6$ \\
11042923\tablenotemark{1} & 0.76 & 0.390 & $984.4 \pm 63.9$ & $0.258 \pm 0.002$ & $276 \pm 11$ \\
11246163 & 0.77 & 0.279 & $1010.6 \pm 181.3$ & $0.164 \pm 0.005$ & $69 \pm 8$ \\
11347875 & 0.86 & 3.455 & $1180.4 \pm 547.1$ & $0.000 \pm 0.011$ & $453 \pm 140$ \\
11604958 & 0.72 & 0.299 & $1068.9 \pm 229.6$ & $0.000 \pm 0.006$ & $41 \pm 5$ \\
11716688 & 0.94 & 0.301 & $1371.3 \pm 2011.1$ & $0.477 \pm 0.004$ & $462 \pm 452$ \\
11825204 & 0.98 & 0.210 & $1336.0 \pm 1253.8$ & $0.414 \pm 0.004$ & $682 \pm 426$ \\
12019674 & 0.76 & 0.355 & $1088.4 \pm 34.7$ & $0.346 \pm 0.001$ & $198 \pm 4$ \\
12055255 & 0.90 & 0.221 & $1093.4 \pm 22.6$ & $0.479 \pm 0.001$ & $297 \pm 4$ \\
12071741 & 0.94 & 0.314 & $939.0 \pm 78.6$ & $0.737 \pm 0.002$ & $473 \pm 26$ \\
12458133 & 0.76 & 0.333 & $1116.4 \pm 396.0$ & $0.484 \pm 0.012$ & $25 \pm 6$ \\
\tableline 
3114667 & 0.52 & 0.889 & $\sim 500$ & \nodata & \nodata \\
2983113 & 0.89 & 0.395 & $\sim 1300$ & \nodata & \nodata \\
4066203 & 0.93 & 0.363 & $\sim 700$ & \nodata & \nodata \\
4074708 & 0.73 & 0.302 & $\sim 1100$ & \nodata & \nodata \\
4904304 & 0.89 & 0.389 & $\sim 1300$ & \nodata & \nodata \\
5282464 & 0.72 & 0.496 & $\sim 1100$ & \nodata & \nodata \\
5956776 & 0.61 & 0.569 & $\sim 700$ & \nodata & \nodata \\
6153219 & 0.62 & 0.530 & $\sim 700$ & \nodata & \nodata \\
6265720 & 0.92 & 0.312 & $\sim 1300$ & \nodata & \nodata \\
6794131 & 0.81 & 1.613 & $\sim 1300$ & \nodata & \nodata \\
7506164 & 0.88 & 0.558 & $\sim 1300$ & \nodata & \nodata \\
7590728 & 0.69 & 0.477 & $\sim 900$ & \nodata & \nodata \\
9544350 & 0.92 & 2.260 & $\sim 1000$ & \nodata & \nodata \\
9776718 & 0.87 & 0.544 & $\sim 800$ & \nodata & \nodata \\
9788457 & 0.60 & 0.963 & $\sim 1000$ & \nodata & \nodata \\
10095469 & 0.60 & 0.678 & $\sim 800$ & \nodata & \nodata \\
10796477 & 0.74 & 0.485 & $\sim 1100$ & \nodata & \nodata \\
2159783 & 0.87 & 0.374 & $\sim 1900$ & \nodata & \nodata \\
3127873 & 0.91 & 0.672 & $\sim 4200$ & \nodata & \nodata \\
3221207 & 0.81 & 0.474 & $\sim 1700$ & \nodata & \nodata \\
3342425 & 0.93 & 0.393 & $\sim 6200$ & \nodata & \nodata \\
3766353 & 0.53 & 2.667 & $\sim 6100$ & \nodata & \nodata \\
3848042 & 0.99 & 0.411 & $\sim 2600$ & \nodata & \nodata \\
3935319 & 0.75 & 0.353 & $\sim 1400$ & \nodata & \nodata \\
3936357\tablenotemark{4} & 0.76 & 0.369 & $\sim 2400$ & \nodata & \nodata \\
4077442 & 0.58 & 0.693 & $\sim 6900$ & \nodata & \nodata \\
4464999 & 0.77 & 0.434 & $\sim 3000$ & \nodata & \nodata \\
4563150 & 0.79 & 0.275 & $\sim 2000$ & \nodata & \nodata \\
4758368 & 0.57 & 3.750 & $\sim 1500$ & \nodata & \nodata \\
4945588 & 0.99 & 1.129 & $\sim 1500$ & \nodata & \nodata \\
5008287 & 0.94 & 0.292 & $\sim 2300$ & \nodata & \nodata \\
5015926 & 0.75 & 0.363 & $\sim 1400$ & \nodata & \nodata \\
5097446 & 0.60 & 1.288 & $\sim 6100$ & \nodata & \nodata \\
5296877 & 0.95 & 0.377 & $\sim 1900$ & \nodata & \nodata \\
5353374 & 0.78 & 0.393 & $\sim 1700$ & \nodata & \nodata \\
5389616 & 0.99 & 0.407 & $\sim 2100$ & \nodata & \nodata \\
5513861\tablenotemark{2} & 0.57 & 1.510 & $\sim 1800$ & \nodata & \nodata \\
5770431 & 0.89 & 0.392 & $\sim 4900$ & \nodata & \nodata \\
5820209 & 0.81 & 0.656 & $\sim 1600$ & \nodata & \nodata \\
5951553 & 0.95 & 0.432 & $\sim 2300$ & \nodata & \nodata \\
6144827 & 0.79 & 0.235 & $\sim 5000$ & \nodata & \nodata \\
6187893 & 0.59 & 0.789 & $\sim 7800$ & \nodata & \nodata \\
6287172 & 0.95 & 0.204 & $\sim 1400$ & \nodata & \nodata \\
6370361 & 0.84 & 0.455 & $\sim 6200$ & \nodata & \nodata \\
7137798 & 0.51 & 2.254 & $\sim 6700$ & \nodata & \nodata \\
7269843 & 0.79 & 0.268 & $\sim 6200$ & \nodata & \nodata \\
7339345 & 0.74 & 0.260 & $\sim 2500$ & \nodata & \nodata \\
7367833 & 0.76 & 0.286 & $\sim 2200$ & \nodata & \nodata \\
7375612\tablenotemark{4} & 0.97 & 0.160 & $\sim 2100$ & \nodata & \nodata \\
7680593 & 0.97 & 0.276 & $\sim 1500$ & \nodata & \nodata \\
7697065 & 0.69 & 0.273 & $\sim 2100$ & \nodata & \nodata \\
7709086 & 0.78 & 0.409 & $\sim 2000$ & \nodata & \nodata \\
7773380 & 0.94 & 0.308 & $\sim 1400$ & \nodata & \nodata \\
7871200 & 0.74 & 0.243 & $\sim 6800$ & \nodata & \nodata \\
7878402 & 0.73 & 0.374 & $\sim 7700$ & \nodata & \nodata \\
8016214 & 0.53 & 3.175 & $\sim 2100$ & \nodata & \nodata \\
8122124\tablenotemark{4} & 1.00 & 0.249 & $\sim 1600$ & \nodata & \nodata \\
8189196 & 0.98 & 2.304 & $\sim 8300$ & \nodata & \nodata \\
8222945 & 0.99 & 0.451 & $\sim 2300$ & \nodata & \nodata \\
8231231 & 0.89 & 0.712 & $\sim 1600$ & \nodata & \nodata \\
8257903 & 0.76 & 0.515 & $\sim 4100$ & \nodata & \nodata \\
8265951 & 0.81 & 0.780 & $\sim 2000$ & \nodata & \nodata \\
8703528 & 0.74 & 0.400 & $\sim 1500$ & \nodata & \nodata \\
8715667 & 0.85 & 0.406 & $\sim 6200$ & \nodata & \nodata \\
8758716\tablenotemark{4} & 1.00 & 0.107 & $\sim 4900$ & \nodata & \nodata \\
9026766 & 0.75 & 0.272 & $\sim 8900$ & \nodata & \nodata \\
9083523 & 0.65 & 0.918 & $\sim 5200$ & \nodata & \nodata \\
9097798 & 0.99 & 0.334 & $\sim 4000$ & \nodata & \nodata \\
9181877 & 0.74 & 0.321 & $\sim 2600$ & \nodata & \nodata \\
9347868 & 0.82 & 0.318 & $\sim 9500$ & \nodata & \nodata \\
9657096 & 0.94 & 2.138 & $\sim 1400$ & \nodata & \nodata \\
9724080 & 0.94 & 1.174 & $\sim 3000$ & \nodata & \nodata \\
9882280 & 0.75 & 0.289 & $\sim 3100$ & \nodata & \nodata \\
9956124 & 0.91 & 0.363 & $\sim 8800$ & \nodata & \nodata \\
10007533 & 0.88 & 0.648 & $\sim 5600$ & \nodata & \nodata \\
10135584 & 0.96 & 0.391 & $\sim 4200$ & \nodata & \nodata \\
10216186 & 0.64 & 0.606 & $\sim 5400$ & \nodata & \nodata \\
10228991 & 0.97 & 2.799 & $\sim 1600$ & \nodata & \nodata \\
10229723 & 0.83 & 0.629 & $\sim 7400$ & \nodata & \nodata \\
10481912 & 0.96 & 0.442 & $\sim 2700$ & \nodata & \nodata \\
10485137 & 0.71 & 0.445 & $\sim 3100$ & \nodata & \nodata \\
10557008 & 0.77 & 0.265 & $\sim 1500$ & \nodata & \nodata \\
10711938 & 0.95 & 0.358 & $\sim 2000$ & \nodata & \nodata \\
11091082 & 0.82 & 0.385 & $\sim 1400$ & \nodata & \nodata \\
11151970 & 0.87 & 0.312 & $\sim 9000$ & \nodata & \nodata \\
11305087 & 0.78 & 0.309 & $\sim 5300$ & \nodata & \nodata \\
11496078 & 0.87 & 0.300 & $\sim 6500$ & \nodata & \nodata \\
11509282 & 0.83 & 0.634 & $\sim 1600$ & \nodata & \nodata \\
11566174 & 0.76 & 0.277 & $\sim 3200$ & \nodata & \nodata \\
11717798 & 0.83 & 0.375 & $\sim 8000$ & \nodata & \nodata \\
11805235 & 0.79 & 0.395 & $\sim 5400$ & \nodata & \nodata \\
11910076 & 0.96 & 0.348 & $\sim 4600$ & \nodata & \nodata \\
12055421 & 0.96 & 0.386 & $\sim 1600$ & \nodata & \nodata \\
12267718 & 0.89 & 0.545 & $\sim 7000$ & \nodata & \nodata \\
12554536 & 0.63 & 0.684 & $\sim 2200$ & \nodata & \nodata \\
\enddata
\tablecomments{This table is divided into three sections based on the period for the third body found by the LTTE fit.  
The first section includes periods from 0-700 days (approximately half the length of the time-baseline of the photometric data used), the second from 700-1400 days (approximately the length of the time baseline), and the third longer than 1400 days.
 In the first section, the fits are based on at least 2 full cycles of the LTTE orbit, the second on 1 full cycle, and the third is a very preliminary fit based on some evidence of curvature in the ETVs. Some binaries with third body signals that fell in the second section did not have \emph{Kepler} data available for the full baseline.  These no longer met the criteria of having a full cycle of data, and so were moved to the beginning of the third section of the table. 
These three different sections are noted in all plots with black, gray, and light gray respectively.}
\tablenotetext{1}{Appears in \citet{Rapp}}
\tablenotetext{2}{Appears in \citet{Gies} as candidate third-body}
\tablenotetext{3}{\citet{KepEB2}}
\tablenotetext{4}{Shows depth variations}
\tablenotetext{5}{Visible tertiary eclipse}
\end{deluxetable}

\end{document}